\documentclass[final]{aipproc}

\layoutstyle{6x9}

\begin{document}

\title{Phase Motion in the Scalar Low-Mass  $\pi^-\pi^+$ Amplitude in 
$D^+ \to \pi^-\pi^+\pi^+$}

\author{Alberto Reis, for the Fermilab E791 Collaboration}{
  address={Centro Brasileiro de Pesquisas F\'{\i}sicas - CBPF - Brazil}
}

\begin{abstract}

 This work is a direct and model-independent
  measurement of the
low mass $\pi^+\pi^-$  phase  motion in the $D^+ \to \pi^-\pi^+\pi^+$ decay.
 The  results show  a strong phase variation,    compatible with
 an isoscalar  $\sigma(500)$ meson.  This result confirms the previous Fermilab
 E791  result which found evidence for the existence of  this scalar
 particle using a full Dalitz-plot analysis.
\end{abstract}

\maketitle

\section{Introduction}

  Charm meson decays are a natural place for studying light scalar mesons.
These decays are a clean environment: well defined initial state, very small
non-resonant component and large coupling to scalars. In addition, D decays can
provide insight into the quark content of these controversial particles, 
since the bulk of the hadronic
width comes from modes for which there is a W-radiation amplitude.

  Studies of light scalars using charm decays started a
few years ago, when E791\cite{e7910} presented results of Dalitz-plot analyses of the 
decays $D^+_s,D^+\to \pi^-\pi^+\pi^+$\cite{e7911,e7912} and $D^+ \to K^-\pi^+\pi^+$ 
\cite{e7913}. In particular, the analysis of the decays $D^+\to \pi^-\pi^+\pi^+$
and $D^+ \to K^-\pi^+\pi^+$ showed evidence for the $\sigma$ and $\kappa$ mesons. 
In the case of $D^+\to \pi^-\pi^+\pi^+$, the $\sigma$ appears as an accumulation of
signal events in low  $\pi^+\pi^-$ mass. The E791 data could only be described by an
amplitude having an $s$ dependent phase ($s$ being the $\pi^- \pi^+$ mass 
squared).

Several years later, the same effect - an accumulation of signal events in low  
$\pi^+\pi^-$ mass - was also
 observed in other decays from different experiments, for, instance,
$D^0 \to \bar K^0 \pi^+\pi^-$ from CLEO\cite{cleo} and Belle\cite{belle}, and
$J/\psi \to \omega \pi^+\pi^-$ from BES\cite{bes}. Again, the description of these data
requires amplitudes with $s$-dependent phase.

Since E791 publication of the $D^+_s,D^+\to \pi^-\pi^+\pi^+$ results, two kinds of
criticisms have been made: the quoted values of both $\sigma$ and $\kappa$ parameters 
are not correct because the simple Breit-Wigner formula is inadequate to describe
broad scalars, especially when near the threshold; the other criticism is that 
one should not claim the existence of a resonance without showing the phase motion of
the corresponding amplitude. 

  The Breit-Wigner formula, although being in this case only a naive approximation, 
has the key ingredient: an $s$-dependent phase.
On the other hand, there is no single agreed
way to treat broad scalars near threshold. The $\sigma$ parameters depend strongly
on the assumed functional form of its line shape.

  In this work\cite{e7914} the second criticism is addressed. The phase 
variation of the  low-mass $\pi^+\pi^-$ amplitude is extracted in a model 
independent way.

\section{The amplitude difference method}

In Dalitz-plot analysis, resonant amplitudes are complex functions written in the 
general form
${\mathcal A} = f(s) e^{i\delta(s)}$. Non-resonant amplitudes, in contrast, have
$ \delta(s)=$ constant. If two resonant amplitudes cross in some region of the 
Dalitz-plot,
they will interfere. The interference pattern in this crossing region
depends on the s-dependent phases of
both resonances. One can extract the phase motion of an unknown amplitude that
crosses  a well known resonance provided that:

\begin{itemize}

\item the contribution of other amplitudes is negligible in the 
crossing region between  the amplitude under 
study and the known resonance;

\item the integrated amplitude of the known resonance is symmetric 
with respect to an effective mass squared ($m^2_{eff}$). 
 
\end{itemize}

The second condition ensures that, by comparing the amplitude below and above 
$m^2_{eff}$, we end up with an expression that involves only the 
desired phase $\delta(s_{13})$. Then, we can write the   
approximate amplitude of this phase space region in a simple way,

\begin{equation} 
 {\cal A}(s_{12},s_{13}) \simeq ~a_{R} ~{\cal BW}(s_{12}) ~{\cal M}(s_{12},s_{13}) + 
 ~a_s/(p^*/\sqrt s_{13}) ~sin \delta (s_{13}) ~e^{i(\delta(s_{13})+\gamma)}
\end{equation} 

\noindent where $\gamma$ is the overall relative final state
interaction (FSI) phase difference between the two amplitudes, $a_R$ and $a_s$
are respectively the real magnitudes of the known resonance and the under-study
complex amplitude, $sin \delta (s_{13}) e^{i\delta(s_{13})}$ represents
the most general amplitude for a two-body elastic scattering; 
$p^*/\sqrt s_{13}$ is a phase space factor to make this description 
compatible with $\pi \pi$ scattering and  
${\cal M}(s_{12},s_{13})$, ${\cal BW}(s_{12})$ are the angular function and 
Breit-Wigner for the known resonance, respectively.

The quantity  
$\Delta\mid { \cal A}( s_{13})\mid^2 \equiv \mid {\cal A}(m_{eff}^2 + \epsilon, s_{13}) 
\mid^2 -\mid {\cal A}( m_{eff}^2 - \epsilon, s_{13}) \mid^2$, which is the difference of 
the amplitudes squared after integration over 
$s_{12}$, is computed in bins of  $s_{13}$. It takes the form

\begin{equation}
\Delta\mid { \cal A}( s_{13})\mid^2 = {{- 4 a_s a_R/(p^*/\sqrt s_{13}) \epsilon m_0 \Gamma_0 
\over\epsilon^2 + m_0^2\Gamma_0^2} ~(sin(2 \delta(s_{13})+ \gamma) - sin \gamma)}
{\cal M}(s_{13})/(p^*/\sqrt s_{13}) 
\end{equation}
\vspace{.5cm}

If $\delta( s_{13})$  is an analytical function of $s_{13}$, then there will be
maximum and minimum values of $\Delta\mid { \cal A}( s_{13})\mid^2$, which we can
use to determine both the constant term in the above equation and the phase $\gamma$,

\begin{equation}
{- 4 a_s a_R/(p^*/\sqrt s_{13}) \epsilon m_0 \Gamma_0 
\over\epsilon^2 + m_0^2\Gamma_0^2} 
\equiv {\cal C} = (\Delta\mid{\cal A}'\mid^2_{max}-\Delta\mid{\cal A}'\mid^2_{min})/2 
\end{equation}

\begin{equation}
\gamma = sin^{-1} ({\Delta \mid { \cal A}'\mid^2_{max} + \Delta\mid { \cal A}'\mid^2_{min}
\over \Delta \mid {\cal A}'\mid^2_{min}  -\Delta\mid  { \cal A}'\mid^2_{max} } )
\end{equation}

\noindent where ${\cal A}' \equiv {\cal A} /({\cal M} \sqrt s_{13}/p^*)$.
\vspace{.3cm}

Finally, considering $\delta( s_{13})$ an increasing  function of $s_{13}$, we have 

\begin{equation} 
\delta (s_{13}) = {1\over 2} ( sin^{-1} ({1 \over { \cal C}} 
\Delta\mid { \cal A}'( s_{13})\mid^2 + sin(\gamma)) - \gamma )
\end{equation}
\vspace{.3cm}

This is, in essence, the idea of the amplitude 
difference method \cite{igju}. The method was shown to work in a "calibration" 
exercise using the $f_0(980)$
resonance in the $D_s^+ \to \pi^- \pi^+ \pi^+$ decay \cite{utica}. In this case we
have the $f_0(980)$ contribution in both $s_{12}$ and $s_{13}$ axes. We were able
to get a phase motion $\delta( s_{13})$ compatible with the $f_0(980)$ using the
$f_0(980)$ in $s_{12}$.

\section{Phase motion of scalar low-mass  $\pi^-\pi^+$ amplitude}

The folded Dalitz-plot distribution of the $D^+ \to \pi^- \pi^+ \pi^+$ 
decay is shown Fig. 2. 
The horizontal  and vertical axes are the squares of the $ \pi^+ \pi^- $ 
invariant mass high ($s_{12}$) and low ($s_{13}$) combinations.

\begin{figure}
  \includegraphics[height=.3\textheight]{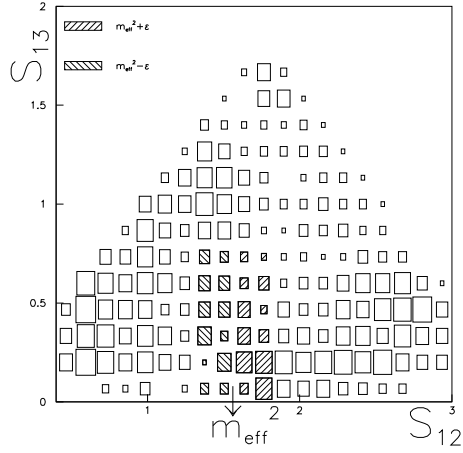}
  \caption{$ D^+ \to \pi^-\pi^+\pi^+ $ Dalitz-plot folded distribution. 
The events used by the AD analysis are in the hatched region. The size of the
area of each bin in the plot corresponds to the number of events in that bin.}
\end{figure}

To study the low mass region in $s_{13}$, there are  
three possible well known resonances in $s_{12}$ to act as a probe  
in this decay: $\rho(770)$, $f_0(980)$ and 
$f_2(1270)$. Figure 1 shows that the $\rho(770)$ and  $f_0(980)$ are located 
in  regions where other amplitudes can not be considered negligible.
On the other hand, the tensor $f_2(1270)$, $m_0^2 = 1.61$ GeV$^2/c^4$,
is placed where the $\rho(770)$, in the 
crossed channel reaches a minimum due to its decay angular distribution.

With the proper choice of $m_{eff}^2 = 1.535$ GeV$^2/c^4$, 
the integral over $s_{12}$ of the $f_2(1270)$ amplitude squared is symmetrical:
the number of events 
 between  $ m_{eff}^2$ and $ m_{eff}^2 + \epsilon$ ($\epsilon = 0.26$ GeV$^2/c^4$) 
 is equal to the number of events between  
 $ m_{eff}^2$ and $ m_{eff}^2 - \epsilon$. 
Moreover, in this region there is no significant
contribution other than the $\pi \pi $  complex amplitude under study in $s_{13}$ 
(the amount of $\rho(770)$ within this mass region was estimated to be 
$\sim$5\%). The choice of the $f_2(1270)$ as the analyser amplitude satisfies the
necessary conditions for the amplitude difference method.

The  acceptance and the background must be similar between
 $ m_{eff}^2$ and $ m_{eff}^2 + \epsilon$ and
$ m_{eff}^2$ and $ m_{eff}^2 - \epsilon$, otherwise there would be
biases in $\delta(s_{13})$. Monte Carlo simulations show that the acceptance is
nearly uniform in this region. The background in this region comes mostly from
random combinations of three pions, and it is also uniformly distributed.
Since we are subtracting  two similar distributions, we considered the
background only in the size of the statistical error.
 
The $f_2(1270)$ angular function has a zero at about $s_{13} \simeq$0.48 GeV$^2/c^4$.
This means a singularity in ${\cal A}'$. This singularity is handled in the following
way. The data is divided into ten $s_{13}$ bins. The binning is such  that 
the singularity is placed in the middle of one bin. Doing  this,  we isolate the
singularity in a single bin (bin 6) and discard its further use in the analysis.

The values of $\gamma$ and ${\cal C}$ were obtained solving 
Equations 3 and 4. The value of
the phase $\gamma$ from the amplitude difference method is in  agreement 
with that of the full Dalitz-plot analysis,
$\gamma_{AD} = 2.78 \pm 0.38 \pm 0.40$ and $\gamma_{Dalitz} = 2.59 \pm 0.19$\cite{e7912}.

Having $\gamma$ and ${\cal C}$, the phase $\delta(s_{13})$ is obtained for each
$s_{13}$ bin using Equation 5. There are  ambiguities  
arising from  the $sin^{-1}$ operation, so  $\delta(s_{13})$ was
determined with the assumption that  
the phase difference starts at zero at threshold and is an  increasing, 
monotonic, smooth function  of  $ s_{13}$.

The phase motion of the low $\pi^+\pi^-$ mass amplitude,
including systematic and statistical errors, is shown in Fig. 2. In spite of the 
limited statistics, a strong phase variation is clearly observed.
Starting from zero at the threshold  the phase varies by about 180$^0$ 
and saturates at around $s_{13} = 0.6$ GeV$^2/c^4$. This is the expected behavior of
resonance. The observed phase motion supports the interpretation of the $\sigma(500)$
as a true resonance. A constant or slowly varying phase would disfavor this
interpretation.
With more statistics we could have more bins and the pole position could be inferred.  
Fig. 2 shows also the phase motion of the simple Breit-Wigner (solid line)
used in \cite{e7912}. Even considering the Breit-Wigner as a naive approximation, there
is a qualitative agreement between its phase motion and the directly extracted
$\delta(s_{13})$: the description of the data requires an amplitude with a strong
phase variation.

\begin{figure}
  \includegraphics[height=.3\textheight]{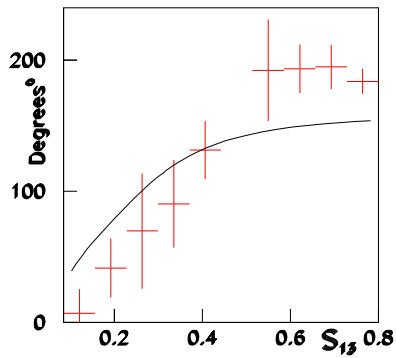}
  \caption{Phase motion vs $s_{13}$, with  errors shown (systematic and
statistical in quadrature). The continuous line is  the
Breit-Wigner phase motion, with the E791 parameters
for the  $\sigma(500)$.}
\end{figure}

\section{Conclusions}

A direct, model-independent measurement of 
the phase motion of the low mass $\pi^+\pi^-$ scalar amplitude was discussed.
Using  the well 
known  $f_2(1270)$ tensor meson in the crossing  channel as the base 
resonance, from the $D^+ \to \pi^-\pi^+\pi^+$ decay, the $\delta( s_{13})$ phase 
motion was extracted. 
We obtain a $\delta( s_{13})$ variation of about 180$^0$, which is the expected
behavior of a resonant amplitude. This result supports the interpretation of the
$\sigma(500)$ as a true resonance, in agreement with the conclusions from
our previous  analysis of full $D^+ \to \pi^- \pi^+ \pi^+$ Dalitz-plot. We could not
extract the $\sigma$ pole position with this method due to the limited statistics.
The measurement of the correct $\sigma$ pole position from a full Dalitz-plot analysis
needs the correct functional form (theorists should agree on what it is). 
Whether the $\sigma$ pole is the same in charm decay and scattering remains an open
question.

\bibliographystyle{aipproc}   

\end{document}